\documentclass[
    ,final            
  ,numberedheadings 
  ]
  {aipproc}

\layoutstyle{8x11single}

\usepackage{amssymb,amsmath,wasysym}
\usepackage[usenames,dvips]{color}

\newcommand{\msun}{M_\odot}

\newcommand{\Ebind}{E_\mathrm{bind}}
\newcommand{\lam}{\lambda_\mathrm{env}}
\newcommand{\alp}{\alpha_\mathrm{CE}}
\newcommand{\eg}{\emph{e.g.}}

\begin{document}

\title{Masses and envelope binding energies of primary stars\\at the onset of a common envelope}


\author{Marc van der Sluys}{
  address={University of Alberta, Canada;~ CITA National Fellow;~ now at R.U.\ Nijmegen, The Netherlands}
}

\author{Michael Politano}{
  address={Marquette University, Milwaukee, Wi, U.S.A.}
}

\author{Ronald E.\ Taam}{
  address={Northwestern University;~ Academia Sinica Institute of Astronomy \& Astrophysics/TIARA, Taipei, Taiwan}
}

\begin{abstract}
  We present basic properties of primary stars that initiate a common envelope (CE) in a binary, while on the giant branch.  We use the
  population-synthesis code described in \citet{2010ApJ...720.1752P} and follow the evolution of a population
  of binary stars up to the point where the primary fills its Roche lobe and initiates a CE.  We then collect the properties of each
  system, in particular the donor mass and the binding energy of the donor's envelope, which are important for the treatment of a CE. 
  We find that for most CEs, the donor mass is sufficiently low to define the core-envelope boundary reasonably well.
  We compute the \emph{envelope-structure parameter} $\lam$ from the binding energy and compare its distribution to typical
  assumptions that are made in population-synthesis codes.  We conclude that $\lam$ varies appreciably and that the assumption of a
  constant value for this parameter results in typical errors of 20--50\%.  In addition, such an assumption may well result in
  the implicit assumption of unintended and/or unphysical values for the CE parameter $\alp$.  Finally, we discuss accurate 
  existing analytic fits for the envelope binding energy, which make these oversimplified assumptions for $\lam$, and the use of 
  $\lam$ in general, unnecessary.
\end{abstract}

\maketitle


\section{Introduction}
\label{sec:intro}

Common envelopes (CEs) \citep{1976IAUS...73...75P,1984ApJ...277..355W,2000ARA&A..38..113T} constitute an important phase in the 
evolution of many binaries and are used to explain the formation of large numbers of observed compact binaries, such as X-ray 
binaries, cataclysmic variables and double white dwarfs.
A CE is assumed to be initiated when a giant star with a deep convective envelope in orbit with a more compact, sufficiently
low-mass companion fills its Roche lobe and the ensuing mass transfer is unstable \citep{1996ApJ...471..366R}.
The result is a fast expanding envelope which can quickly engulf the companion star.  The core of the donor star and the companion
star orbit each other inside this CE, and the friction and torques lead to the spiral-in of the orbit, resulting
in a compact binary or a merger.  The energy generated by the orbital decay is assumed to heat the CE and eventually expel it
from the system on a time scale much shorter than the evolutionary timescales of stars ($\lesssim\,10^3$\,yr), leaving the
secondary star unaffected \citep{2000ARA&A..38..113T}.

Because of the large range of length scales involved, detailed modeling of the CE process is computationally expensive, and
while three-dimensional models can follow about the first month of the process, the outcome of CEs cannot yet be predicted 
\citep[\eg,][]{2008ApJ...672L..41R}.  
In practise, therefore, a cartoonish approach is often used to determine the orbital shrinkage during the spiral-in, 
where the energy needed to unbind the envelope is assumed to be dissipated from the orbit \citep{1979IAUS...83..401T}.  
Hence, the decrease in orbital energy during the CE is related
to the binding energy of the convective envelope of the donor star at the onset of the CE \citep{1984ApJ...277..355W}:
\begin{equation}
  E_\mathrm{bind} = -\alp \left(\frac{G M_\mathrm{1,c} M_2}{2 a_\mathrm{f}} - \frac{G M_1 M_2}{2 a_\mathrm{i}}\right).
  \label{eq:ce}
\end{equation}
In this equation, $E_\mathrm{bind}$ is the binding energy of the donor's convective envelope, $M_1$ is the mass of the
donor at the onset of the CE, $M_\mathrm{1,c}$ is the helium-core mass of the donor and the mass of its remnant if the binary
survives the CE, $M_2$ is the unchanged companion mass, $a_\mathrm{i,f}$ are the initial and final orbital separation, respectively, 
and $\alpha_\mathrm{CE}$ is the efficiency with which the orbital energy is used to expel the envelope.

We developed a population-synthesis code that is tailored to study the first CE event in a binary \citep{2010ApJ...720.1752P}.  
In addition, this code allows us to study the post-CE evolution of the merger remnant or compact binary. 
In these proceedings, we briefly describe our population-synthesis code in Sect.\,\ref{sec:code}.  In Section~\ref{sec:mergers}
we summarise our work on the present-day population of merger remnants that were formed through a CE, which we 
presented in Mykonos.  The bulk of this paper, Section~\ref{sec:donors}, consists of new results, where we present some basic 
properties of donor stars at the onset of a CE.  In Section~\ref{sec:discussion} we discuss these results and present our 
conclusions.

\section{Population-synthesis code}
\label{sec:code}

We developed a population-synthesis code that is specifically targeted at binaries in which the primary star fills its Roche lobe
and causes a CE event \citep{2010ApJ...720.1752P}.  For each binary formed at a random time during the last $10^{10}$ years, 
the code follows the evolution, including effects such as stellar wind and tides, and determines whether the stars merge 
during the CE.  It can then follow the evolution of the merger remnants and create a population of merger products at the 
present epoch.  As input for the code we use a grid of 32 low-mass/brown-dwarf models \citep{2000ApJ...542..464C,2003A&A...402..701B} 
and 116 detailed stellar-evolution models, computed with the binary-evolution code \texttt{ev}\footnote{The 
  current version of \texttt{ev} is obtainable on request from \texttt{eggleton1@llnl.gov}, along with data files and a user manual.}, 
developed by Eggleton \citep[][and references therein]{1971MNRAS.151..351E, 1972MNRAS.156..361E,2005ApJ...629.1055Y} and updated as 
described in \citet{1995MNRAS.274..964P}.  We define the helium-core mass $M_\mathrm{c}$ as the inner region of the star where 
the hydrogen abundance lies below 10\% ($X<0.1$).  The detailed stellar-structure models allow us to compute the envelope binding 
energy accurately by integrating the stellar structure over the mass coordinate from the core-envelope boundary $M_\mathrm{c}$ to 
the stellar surface $M_\mathrm{s}$:
\begin{equation}
  \Ebind = \int_{M_\mathrm{c}}^{M_\mathrm{s}} E_\mathrm{int}(m) - \frac{G m}{r(m)}\, \mathrm{d}m,
  \label{eq:Ebind}
\end{equation}
where internal energy $E_\mathrm{int}$ contains the thermal and radiation energy of the gas, but not its recombination energy.  
More details of the code as used for these models are described in \citet{2010ApJ...720.1752P}.

\section{The present-day population of CE mergers}
\label{sec:mergers}

At the conference, we presented the results of our study of a population of 25,000 present-day merger remnants,
descendant from an initial population of $10^7$ ZAMS binaries with component masses up to $10\,\msun$ each.
These results have recently been published in \citet{2010ApJ...720.1752P}.  We will therefore only reiterate the most
important conclusions here, and refer the reader to that paper for details.  The main conclusions of our merger study are:
\begin{itemize}
  \item Between 0.24\% and 0.33\% of the initial ZAMS binaries are visible today as non-degenerate merger products.
  \item Present-day merger remnants constitute of 37\% RGB stars, 57\% HB stars, and 6\% AGB stars.
  \item RGB stars are under-represented w.r.t.\ HB+AGB stars in a merger population, compared to normal single stars.
  \item The median rotational velocity of the merger population is $16.2$\,km\,s$^{-1}$, compared to $2.3$\,km\,s$^{-1}$ for single stars.
\end{itemize}

\section{Properties of donor stars at the onset of a CE}
\label{sec:donors}

\begin{table}
  \begin{tabular}{lrrr}
    \hline
    & \tablehead{1}{r}{b}{RGB}
    & \tablehead{1}{r}{b}{AGB}
    & \tablehead{1}{r}{b}{Total}   \\
    \hline
    Survivors & ~16.1\% & ~31.8\% &  ~48.0\%  \\
    Mergers   & ~45.4\% &  ~6.6\% &  ~52.0\% \\
    Total     & ~61.5\% & ~38.5\% & ~100.0\% \\
    \hline
  \end{tabular}
  \caption{
    Percentages of CEs that are initiated on the RGB or AGB, and either survive as a binary or merge,
    using 165,007 CEs.
    \label{tab:survmerg}
  }
\end{table}

In this main section of these proceedings, we will discuss the properties of the primary stars in our model binaries,
at the moment they fill their Roche lobes and initiate a CE phase.  The CE may result in either the survival
of the binary system, or the merger of the two binary components.  
We consider a population of $10^6$ ZAMS binaries with maximum component masses of $20\,\msun$ each and a uniform distribution 
of mass ratios, resulting in 165,007 CEs that occur when the donor star is on the RGB or AGB.  Table~\ref{tab:survmerg}
lists the fractions of these CEs that lead to survival of the binary or to merger for both RGB and AGB primaries. 
While \citet{2010ApJ...720.1752P} found that for CEs that lead to merger, 
the vast majority (87\%) of CEs occur when the primary is on the RGB rather than AGB, we see that this majority 
drops to $\sim62\%$ when both mergers and survivors 
are taken into account.  Also, as one would intuitively expect because of the 
smaller orbital separations and higher binding energies, most ($74\%$) RGB CEs lead to merger, whereas most (83\%) 
CEs initiated when the primary is on the AGB allow the binary to survive.

\subsection{Donor masses}
\label{sec:mass}

While the combination of Eqs.\,\ref{eq:ce} and \ref{eq:Ebind} seems to fully determine the simplified CE treatment 
based on energy conservation for a given efficiency factor $\alp$, the situation is not quite that simple.  One of
the major obstructions is the definition of the core-envelope boundary, and hence the helium-core mass $M_\mathrm{c}$,
which is needed in Eq.\,\ref{eq:Ebind} to compute the binding energy of the envelope. \citet{2000A&A...360.1043D} and
\citet{2001A&A...369..170T} compare five different definitions of the core-envelope boundary, and discard the two extreme
cases, leaving the three definitions in the middle three rows of Table~1 of the latter paper (one of which is identical
to the definition we use: $X<0.1$).  They find that for low-mass stars ($M\lesssim 7-10\,\msun$), where the gradients in the stellar 
structure are steep, the three definitions give similar results.  However, for more-massive stars, the three results 
become very different and the exact definition of the core mass becomes uncertain by more than an order of magnitude 
for a $20\,\msun$ star.  In fact, the exact separation between remnant and ejecta, and hence the mass of the remnant, 
will depend on the response of both the stellar structure and the orbit to the mass loss, and hence will depend on the 
secondary star as well \citep{2010ApJ...719L..28D,Ivanova10}.

\begin{figure}
  \begin{centering}
    \includegraphics[width=0.468\textwidth]{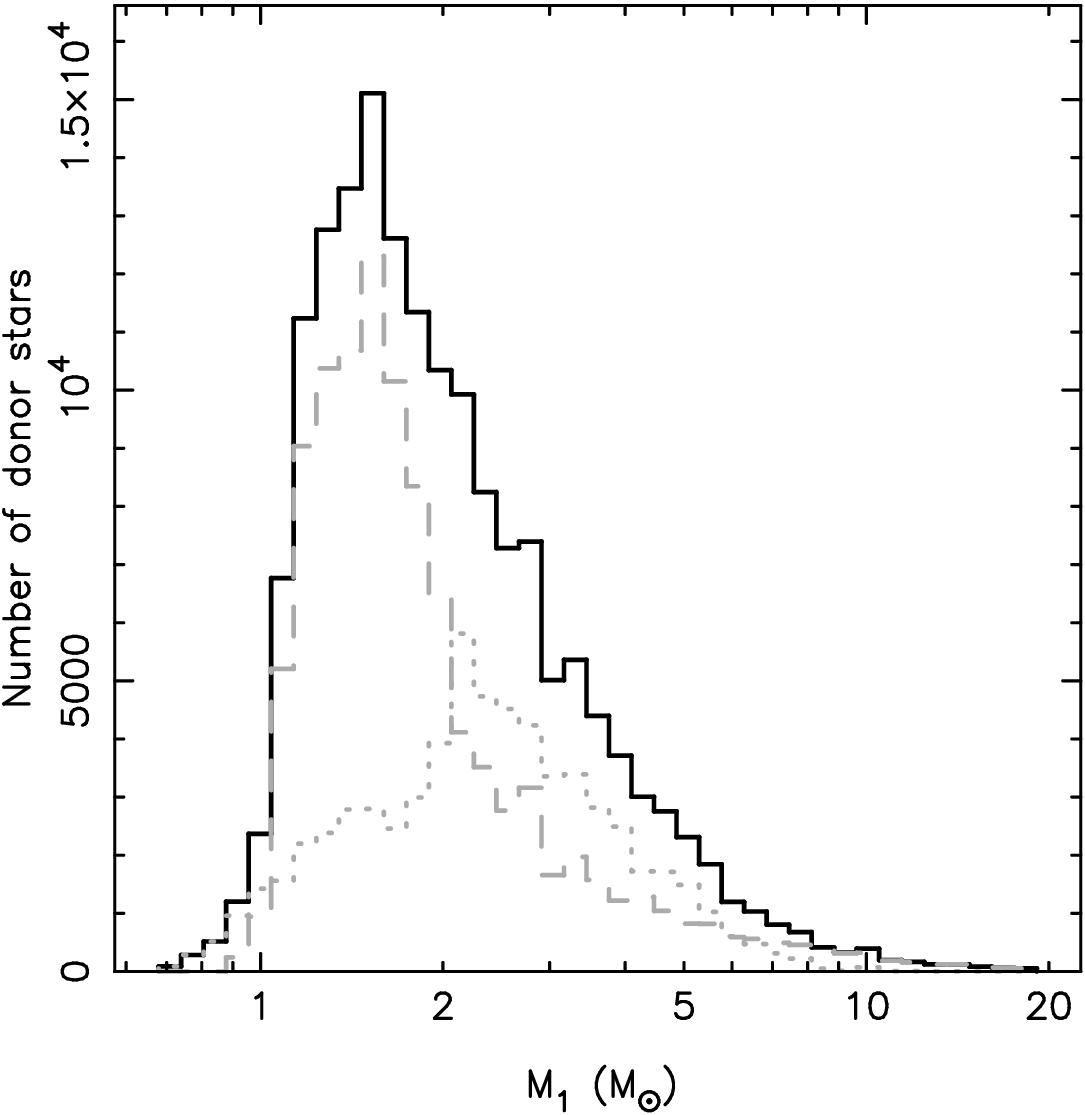}
    
    \hspace*{0.04\textwidth}
    
    \includegraphics[width=0.468\textwidth]{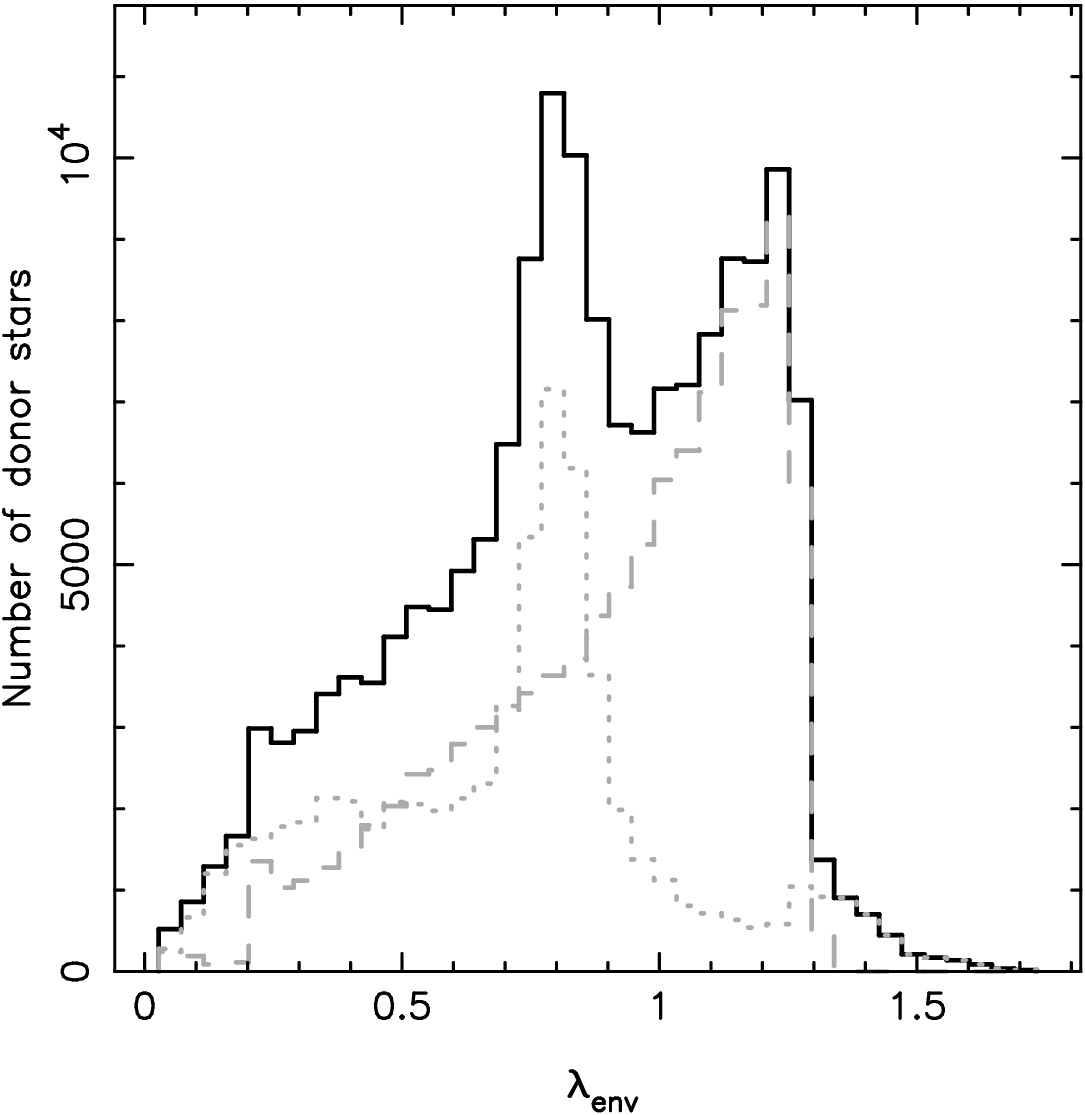}
  \end{centering}
  \caption{
    Distributions of selected properties of the primary stars in our binary population at
    the onset of a CE.
    \textit{Left panel (a):} primary mass $M_1$.
    \textit{Right panel (b):} envelope-structure parameter $\lam$.
    Dashed lines represent donors on the RGB, dotted lines donors on the AGB, and solid
    lines their sum.
    \label{fig:m1_lambda}
  }
\end{figure}

Figure~\ref{fig:m1_lambda}a shows a histogram of donor masses (influenced by wind mass loss) at the onset of their CE
for all donor stars, as well as the separate contributions from RGB and AGB stars.  The median mass of all donors is $1.8\,\msun$ while the RGB
donors contribute more to the low-mass part (median mass $1.6\,\msun$) and the AGB stars more to the high-mass end
(median mass $2.3\,\msun$).  We note that for 98.1\% of the CEs in our models, the donor mass is lower than $7\,\msun$, while
for $99.4\%$ of the CEs, $M_1 < 10\,\msun$.  
This corresponds to $\sim 97.3\%$ and $\sim 98.6\%$ of all CEs respectively when allowing primary masses up to $\sim 200\,\msun$, 
assuming the same fraction of binaries undergo a CE.
This suggests that for most \emph{instances} of a CE, the core mass should be relatively well defined.  However, when 
studying populations of massive stars, \eg\ HMXBs, this is of course no longer true.

\subsection{The envelope-structure parameter $\lam$}
\label{sec:lambda}

In population-synthesis codes, where the evolution of millions of binaries must be computed, detailed stellar-structure models
are computationally too expensive and hence have typically not been used.  
As a consequence, the binding energy of the donor's convective envelope
cannot be computed exactly, and is often approximated using the so-called \emph{envelope-structure parameter} $\lam$ 
\citep{1987A&A...183...47D}, defined by
\begin{equation}
  \Ebind = \frac{G M_1 M_\mathrm{1,env}}{R_1 \lam},
  \label{eq:lambda}
\end{equation}
where $M_\mathrm{1,env} \equiv M_1 - M_{\mathrm{c},1}$ is the mass of the donor's envelope and $R_1$ the donor's radius.  The value for this parameter is different
for different stars and varies through the evolution of a star.  This behaviour is presented in \citet{2000A&A...360.1043D} for 
high-mass stars, and in \citet{2006A&A...460..209V} for lower-mass stars.  Comparison of Eqs.\,\ref{eq:ce} and \ref{eq:lambda} shows that 
the uncertainty in $\alp$ and $\lam$ can be combined
in their product $\alp \lam$.  Hence, as an alternative to assuming a constant value for $\lam$, in many population-synthesis studies a constant 
value for $\alp \lam$ is used.  For example, \citet{2000A&A...360.1011N} and \citet{2002MNRAS.329..897H} 
choose $\lam = 0.5$, while \citet{2008ApJS..174..223B} assume $\alpha_\mathrm{CE} \lam = 0.5, 1.0$.
In our models, the actual binding energy is known through Eq.\,\ref{eq:Ebind}, and therefore we can use Eq.\,\ref{eq:lambda} to compute the 
correct value for $\lam$ for each donor star that initiates a CE.

We present the distribution of $\lam$ for our data set of $165,007$ CEs in Fig.\,\ref{fig:m1_lambda}b.  The distribution is double peaked,
where the peak below $\lam=1$ is caused by AGB donors and the peak above that value by RGB stars.  The range of values for 
$\lam$ is $0.027-1.73$ and the median value of the total distribution is $0.86$, whereas the medians for the RGB and AGB sub-populations 
are $1.00$ and $0.75$ respectively.  Using this distribution, we can derive that the median relative error made when assuming 
$\lam=0.5$ is $47\%$, while that value is $22\%$ for the assumption $\lam=1.0$.  In both cases the extreme errors exceed an order
of magnitude.

\begin{figure}
  \begin{centering}
    \includegraphics[width=0.75\textwidth]{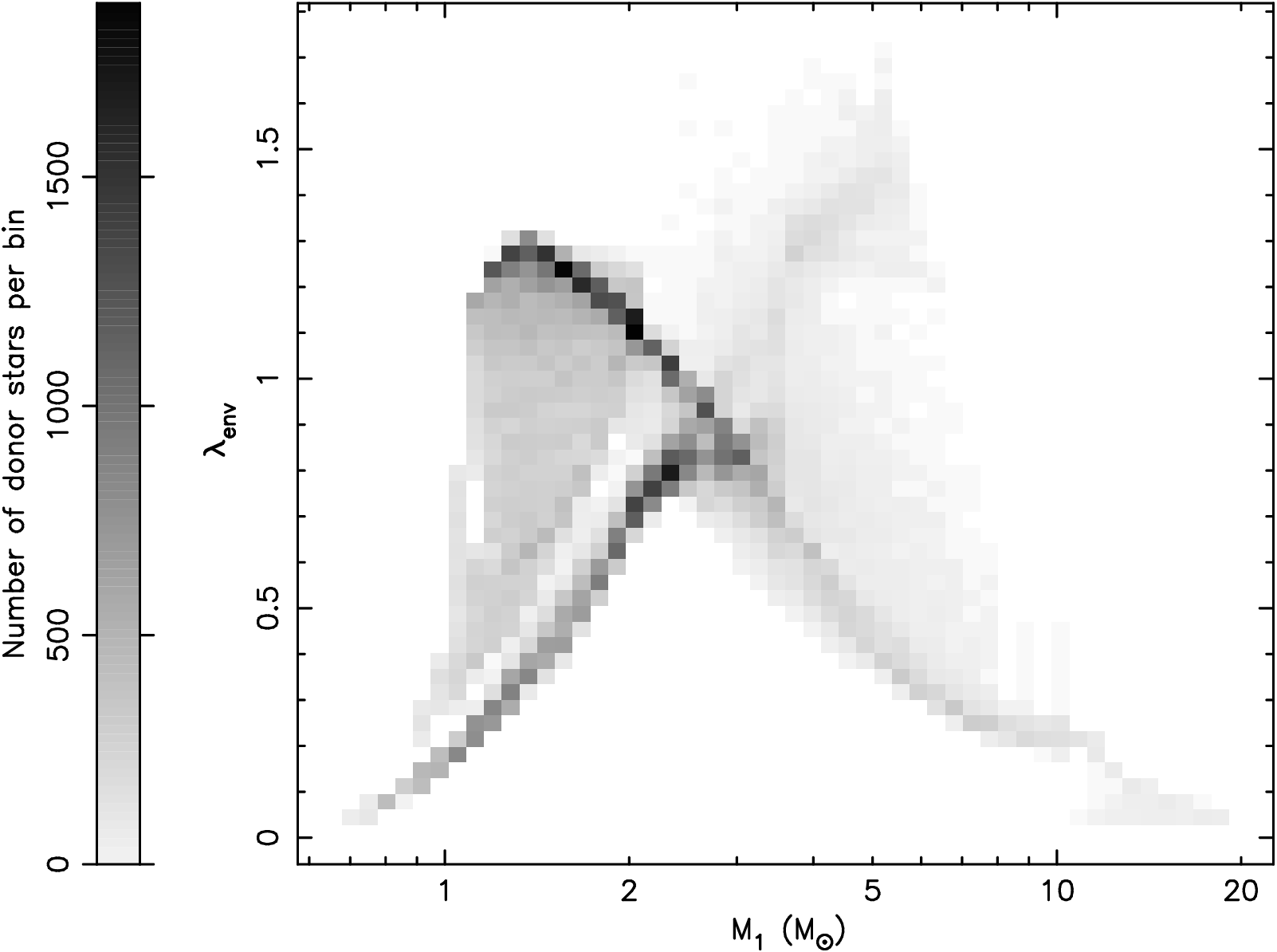}
  \end{centering}
  \caption{
    Two-dimensional histogram displaying $\lam$ versus $M_1$ for primary stars at the onset of a CE.
    Darker pixels indicate more donors in the bin, as indicated by the legend at the left.
    See Sect.\,\ref{sec:lambda} for more details. Note that the pattern looks like a $\lambda$ $\smiley$
    \label{fig:m1-lambda}
  }
\end{figure}

Figure~\ref{fig:m1-lambda} shows a two-dimensional histogram of the $M_1-\lam$ space.  The arm in the $\chi$-shape that runs from 
the upper left to the lower right, as well as the ``slab'' that extends down from the upper-left part, is formed by donor stars on 
the RGB.  The other arm, running from the lower left to the upper right, and the thinly populated area between the two arms 
at $M_1 \gtrsim 3.5\,\msun$,
are formed by stars that initiate a CE on the AGB.  Hence we find that for RGB stars, lower-mass donors have higher values for 
$\lam$, while for AGB donors, this trend is reversed.

\subsubsection{Implications for $\alp$}
\label{sec:alpha}

\begin{figure}
  \begin{centering}
    \includegraphics[width=0.48\textwidth]{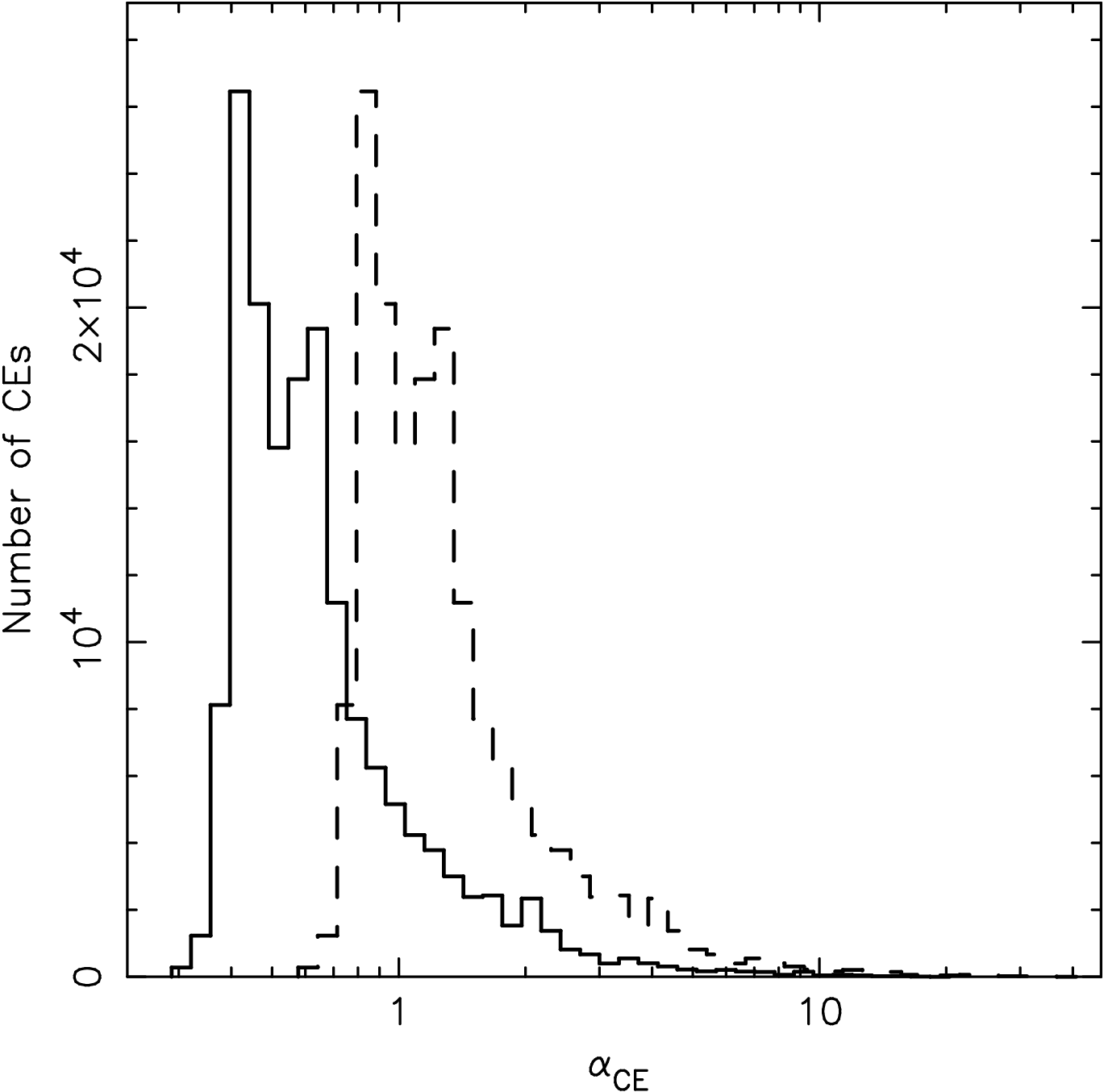}
  \end{centering}
  \caption{
    Distributions of the implicitly assumed common-envelope efficiency parameter $\alp$ when using the premise $\alp \lam=0.5$ (solid line)
    or $\alp \lam = 1.0$ (dashed line).
    \label{fig:alphas}
  }
\end{figure}

Interestingly, we find that for $64.2\%$ of our CE donors the envelope-structure parameter is smaller than unity 
and for $16.3\%$ it is even smaller than 0.5.  
This means that the assumption $\alp \lam = 1.0$ implies $\alp > 1$ 
for most CEs, while the premise 
$\alp \lam = 0.5$ (including \eg\ $\alp=1.0; \lam=0.5$) still gives values for the CE efficiency factor larger than unity for 
about one in six CEs.
Figure~\ref{fig:alphas} displays the distributions of the implicitly assumed values for $\alp$ when using the premises 
$\alp \lam = 0.5$ or $\alp \lam = 1.0$.  We find that $0.29 \lesssim \alp \lesssim 18.3$ for the first assumption (the limiting
values of this range are twice as high for the second assumption), indicating that a few extreme cases are quite unphysical 
indeed.  From Fig.\,\ref{fig:m1-lambda} and its description in Sect.\,\ref{sec:lambda}, we find that extremely high values of $\alp$
(when $\lam$ is low) occur on the AGB for low-mass stars and on the RGB for high-mass stars.

\section{Discussion and conclusions}
\label{sec:discussion}

In the previous section, we argued that although the exact core-envelope boundary is difficult to ascertain for massive stars,
this will affect only $\sim 3\%$ or fewer of CEs, due to the initial-mass function.  On the other hand, for the study of high-mass
compact binaries, this effect may be quite important and even dominate the uncertainty in wind mass loss of massive stars
for the determination of the envelope binding energy (see the discussion in \citet{Loveridge}).

In addition, we presented realistic values for the envelope-structure parameter $\lam$, defined in Eq.\,\ref{eq:lambda},
for stars that initiate a CE.  We find that the actual value of this parameter is far from constant, but varies
from star to star, as well as during the lifetime of a star.  Assuming a constant value for either $\lam$ or $\alp \lam$ introduces
errors of typically $\sim 20-50\%$, but the error may be higher than an order of magnitude in some extreme cases.  Furthermore,
such an assumption for $\lam$ implicitly assumes values for $\alp$ which can be very different from the intended value, and
may be quite unphysical, reaching values of $10$ or more.

It is therefore clear that a more realistic way of estimating the binding energy 
of the convective envelope of a giant star can 
significantly reduce much of the uncertainty with which the post-CE orbit can be determined, 
for a given value of $\alp$.  
Fortunately, and not quite coincidentally (at least in one case), two studies have recently been published that offer a solution.
\citet{2010ApJ...716..114X} offer fits for stars of 14 different discrete masses between $1\msun$ and $20\msun$ for $Z=0.001$
and $Z=0.02$.  The fits provide $\lam$ using the mass and radius (in some cases the core mass replaces the radius) of the star,
which can then be used with Eq.\,\ref{eq:lambda} to compute the binding energy.  Their expressions are simple and easy to
implement, but unfortunately they do not provide the accuracy of their fits, so that it is difficult to assess to what extent
the errors discussed above are improved.  

Independently, \citet{Loveridge} provide fits for six different values of the
metallicity between $Z=10^{-4}$ and $Z=0.03$, which directly give the envelope binding energy for giants with any mass
between $0.8\,\msun$ and $100\,\msun$, as a function of $Z$, $M$ and $R$.  They show that the accuracies of their fits are
better than $15\%$ for $90\%$ of their data points for all metallicities and evolutionary stages, and better than $10\%$ 
for $90\%$ of their data points for all cases, except the AGB for the three lowest metallicities in their data.  This is
a significant improvement from the accuracies found in Sect.\,\ref{sec:lambda}, which can be expressed using the notation 
above as better than 
$22\%$ and $47\%$ for the assumptions $\alp\lam = 0.5$ and $\alp\lam = 1.0$ respectively, for only $50\%$ of the data points.  
The fits are more complex in this study, but the fitting coefficients, and example routines that use them to compute the envelope 
binding energy for a number of cases, are available electronically \citep{Loveridge_edata}.  
Thus, we conclude that there is no longer a need to make oversimplified assumptions for $\lam$,
or in fact no longer a need for the parameter $\lam$ itself, since more accurate alternatives exist to approximate the binding 
energy of the convective envelope for any giant star that can initiate a CE.


\begin{theacknowledgments}
  We would like to thank P.P.\ Eggleton for kindly making his binary-evolution code available to us.  
  MvdS acknowledges support from a CITA National Fellowship to the University of Alberta.  
  M.P.\ acknowledges funding from the Wisconsin Space Grant Consortium and NSF grant AST-0607111, sub-award 1-2008, 
  to Marquette University.  
  R.T.\ acknowledges support from NSF AST-0703950 to Northwestern University.
\end{theacknowledgments}



\bibliographystyle{aipproc}   

\bibliography{vdsluys}

\hyphenation{Post-Script Sprin-ger}
\begin{thebibliography}{25}
\expandafter\ifx\csname natexlab\endcsname\relax\def\natexlab#1{#1}\fi
\providecommand{\enquote}[1]{``#1''}
\expandafter\ifx\csname url\endcsname\relax
  \def\url#1{\texttt{#1}}\fi
\expandafter\ifx\csname urlprefix\endcsname\relax\def\urlprefix{URL }\fi
\providecommand{\eprint}[2][]{\url{#2}}

\bibitem[{Politano} et~al.(2010)]{2010ApJ...720.1752P}
M.~{Politano}, M.~{van der Sluys}, R.~E. {Taam}, and B.~{Willems}, \emph{\apj}
  \textbf{720}, 1752--1766 (2010).

\bibitem[{Paczynski}(1976)]{1976IAUS...73...75P}
B.~{Paczynski}, \enquote{{Common Envelope Binaries},} in \emph{Structure and
  Evolution of Close Binary Systems}, edited by {P.~Eggleton, S.~Mitton, \&
  J.~Whelan}, 1976, vol.~73 of \emph{IAU Symposium}, p.~75.

\bibitem[{Webbink}(1984)]{1984ApJ...277..355W}
R.~F. {Webbink}, \emph{\apj} \textbf{277}, 355--360 (1984).

\bibitem[{Taam} and {Sandquist}(2000)]{2000ARA&A..38..113T}
R.~E. {Taam}, and E.~L. {Sandquist}, \emph{\araa} \textbf{38}, 113--141 (2000).

\bibitem[{Rasio} and {Livio}(1996)]{1996ApJ...471..366R}
F.~A. {Rasio}, and M.~{Livio}, \emph{\apj} \textbf{471}, 366--+ (1996).

\bibitem[{Ricker} and {Taam}(2008)]{2008ApJ...672L..41R}
P.~M. {Ricker}, and R.~E. {Taam}, \emph{\apjl} \textbf{672}, L41--L44 (2008).

\bibitem[{Tutukov} and {Yungelson}(1979)]{1979IAUS...83..401T}
A.~{Tutukov}, and L.~{Yungelson}, \enquote{{Evolution of massive common
  envelope binaries and mass loss},} in \emph{Mass Loss and Evolution of O-Type
  Stars}, edited by {P.~S.~Conti \& C.~W.~H.~De Loore}, 1979, vol.~83 of
  \emph{IAU Symposium}, pp. 401--406.

\bibitem[{Chabrier} et~al.(2000)]{2000ApJ...542..464C}
G.~{Chabrier}, I.~{Baraffe}, F.~{Allard}, and P.~{Hauschildt}, \emph{\apj}
  \textbf{542}, 464--472 (2000).

\bibitem[{Baraffe} et~al.(2003)]{2003A&A...402..701B}
I.~{Baraffe}, G.~{Chabrier}, T.~S. {Barman}, F.~{Allard}, and P.~H.
  {Hauschildt}, \emph{\aap} \textbf{402}, 701--712 (2003).

\bibitem[{Eggleton}(1971)]{1971MNRAS.151..351E}
P.~P. {Eggleton}, \emph{\mnras} \textbf{151}, 351 (1971).

\bibitem[{Eggleton}(1972)]{1972MNRAS.156..361E}
P.~P. {Eggleton}, \emph{\mnras} \textbf{156}, 361 (1972).

\bibitem[{Yakut} and {Eggleton}(2005)]{2005ApJ...629.1055Y}
K.~{Yakut}, and P.~P. {Eggleton}, \emph{\apj} \textbf{629}, 1055--1074 (2005).

\bibitem[{Pols} et~al.(1995)]{1995MNRAS.274..964P}
O.~R. {Pols}, C.~A. {Tout}, P.~P. {Eggleton}, and Z.~{Han}, \emph{\mnras}
  \textbf{274}, 964--974 (1995).

\bibitem[{Dewi} and {Tauris}(2000)]{2000A&A...360.1043D}
J.~D.~M. {Dewi}, and T.~M. {Tauris}, \emph{\aap} \textbf{360}, 1043--1051
  (2000).

\bibitem[{Tauris} and {Dewi}(2001)]{2001A&A...369..170T}
T.~M. {Tauris}, and J.~D.~M. {Dewi}, \emph{\aap} \textbf{369}, 170--173 (2001).

\bibitem[{Deloye} and {Taam}(2010)]{2010ApJ...719L..28D}
C.~J. {Deloye}, and R.~E. {Taam}, \emph{\apjl} \textbf{719}, L28--L31 (2010).

\bibitem[{Ivanova}(2010)]{Ivanova10}
N.~{Ivanova}, \emph{\apj\ submitted}  (2010).

\bibitem[{de Kool} et~al.(1987)]{1987A&A...183...47D}
M.~{de Kool}, E.~P.~J. {van den Heuvel}, and E.~{Pylyser}, \emph{\aap}
  \textbf{183}, 47--52 (1987).

\bibitem[{van der Sluys} et~al.(2006)]{2006A&A...460..209V}
M.~V. {van der Sluys}, F.~{Verbunt}, and O.~R. {Pols}, \emph{\aap}
  \textbf{460}, 209--228 (2006).

\bibitem[{Nelemans} et~al.(2000)]{2000A&A...360.1011N}
G.~{Nelemans}, F.~{Verbunt}, L.~R. {Yungelson}, and S.~F. {Portegies Zwart},
  \emph{\aap} \textbf{360}, 1011--1018 (2000).

\bibitem[{Hurley} et~al.(2002)]{2002MNRAS.329..897H}
J.~R. {Hurley}, C.~A. {Tout}, and O.~R. {Pols}, \emph{\mnras} \textbf{329},
  897--928 (2002).

\bibitem[{Belczynski} et~al.(2008)]{2008ApJS..174..223B}
K.~{Belczynski}, V.~{Kalogera}, F.~A. {Rasio}, R.~E. {Taam}, A.~{Zezas},
  T.~{Bulik}, T.~J. {Maccarone}, and N.~{Ivanova}, \emph{\apjs} \textbf{174},
  223--260 (2008).

\bibitem[{Loveridge} et~al.(2010{\natexlab{a}})]{Loveridge}
A.~J. {Loveridge}, M.~V. {van der Sluys}, and V.~{Kalogera}, \emph{\apj\
  submitted}  (2010{\natexlab{a}}), \eprint{arXiv:astro-ph/1009.5400}.

\bibitem[{Xu} and {Li}(2010)]{2010ApJ...716..114X}
X.~{Xu}, and X.~{Li}, \emph{\apj} \textbf{716}, 114--121 (2010).

\bibitem[{Loveridge} et~al.(2010{\natexlab{b}})]{Loveridge_edata}
A.~J. {Loveridge}, M.~V. {van der Sluys}, and V.~{Kalogera}, Electronic tables
  and routines to compute envelope binding energies of giant stars
  (2010{\natexlab{b}}), http:/$\!$/www.astro.ru.nl/$\sim$sluys/be/.

\end{thebibliography}

\IfFileExists{\jobname.bbl}{}
 {\typeout{}
  \typeout{******************************************}
  \typeout{** Please run "bibtex \jobname" to obtain}
  \typeout{** the bibliography and then re-run LaTeX}
  \typeout{** twice to fix the references!}
  \typeout{******************************************}
  \typeout{}
 }

\end{document}